\documentclass[sigconf]{acmart}

\AtBeginDocument{%
  \providecommand\BibTeX{{%
    Bib\TeX}}}

\copyrightyear{2022} 
\acmYear{2022} 
\setcopyright{acmcopyright}\acmConference[ASE '22]{37th IEEE/ACM International Conference on Automated Software Engineering}{October 10--14, 2022}{Rochester, MI, USA}
\acmBooktitle{37th IEEE/ACM International Conference on Automated Software Engineering (ASE '22), October 10--14, 2022, Rochester, MI, USA}
\acmPrice{15.00}
\acmDOI{10.1145/3551349.3559548}
\acmISBN{978-1-4503-9475-8/22/10}

\usepackage{listings}

\definecolor{dkgreen}{rgb}{0,0.6,0}
\definecolor{gray}{rgb}{0.5,0.5,0.5}
\definecolor{mauve}{rgb}{0.58,0,0.82}

\makeatletter

\newcommand\notsotiny{\@setfontsize\notsotiny{6}{6}} 

\makeatother

\hypersetup{
    colorlinks=true,
    linkcolor=blue,
    filecolor=magenta,      
    urlcolor=cyan,
}

\newcommand{\nd}{\vspace{1mm}\noindent}
\begin{document}

\title{Automatic Code Documentation Generation Using GPT-3}

\author{Junaed Younus Khan and Gias Uddin}
\affiliation{%
  \country{DISA Lab, University of Calgary}
}

\begin{abstract}
Source code documentation is an important artifact for efficient software development. Code documentation could greatly benefit from automation since manual documentation is often labouring, resource and time-intensive. In this paper, we employed Codex for automatic code documentation creation. Codex is a GPT-3 based model pre-trained on both natural and programming languages. We find that Codex outperforms existing techniques even with basic settings like one-shot learning (i.e., providing only one example for training). Codex achieves an overall BLEU score of 20.6 for six different programming languages (11.2\% improvement over earlier state-of-the-art techniques). Thus, Codex shows promise and warrants in-depth future studies for automatic code documentation generation to support diverse development tasks.        

\end{abstract}


\maketitle

\section{Introduction}\label{sec:introduction}
In Software Engineering (SE), developers often try to figure out what a specific code unit (e.g., method) does and how to use it \cite{xia2017measuring}. They can do this by reading documentation of source code. Well-written documentation is crucial for effective software development \cite{DeSouza-DocumentationEssentialForSoftwareMaintenance-SIGDOC2005}. The sudden shift to work from home during COVID-19 further showed the needs for good documentation for developers, when the subject matter expert of code base may be absent/unavailable~\cite{Uddin-QualitativeStudyCovid19-EMSE2022}. However, such documentation is costly and time-consuming to create and maintain. Most developers often show unwillingness towards writing documentation as they find it less productive and less rewarding \cite{parnas2011precise, mcburney2017towards}. As a result, manual documentation often becomes problematic or unusable \cite{Robillard-APIsHardtoLearn-IEEESoftware2009a,Aghajani-SoftwareDocIssueUnveiled-ICSE2019,Aghajani-SoftwareDocPractitioner-ICSE2020,Uddin-UsageScenarioDocumentation-TOSEM2021}. Moreover, documentation becomes obsolete over time with continuous modification or update to the system (i.e., code-base) \cite{ibrahim2012relationship,Uddin-HowAPIDocumentationFails-IEEESW2015,Uddin-TemporalApiUsage-ICSE2012}. 

Automated code documentation generation is currently attracting a lot of attention from the researcher community to substitute or complement the manual documentation efforts. Earlier works in this direction mostly focused on template-based \cite{Sridhara-SummaryCommentsJavaClasses-ASE2010, mcburney2014automatic, moreno2013automatic, rai2017method, abid2015using, rastkar2011generating} and information retrieval (IR)-based strategies \cite{haiduc2010supporting, haiduc2010use, eddy2013evaluating, wong2013autocomment, vassallo2014codes, wong2015clocom}. Template-based approaches are limited to the underlying set of templates and so are finite/limited, while similarity measures in IR-based approaches can be erroneous \cite{rai2022review}. Recently, researchers are investigating several learning-based (e.g., deep learning) approaches for documentation generation. For example, CODE-NN, presented by Iyer et al., can generate documentation of C\# and SQL code using LSTM attention networks \cite{iyer2016summarizing}. Allamanis et al. also used an attention neural network for code summarization \cite{allamanis2016convolutional}. Hu et al. developed DeepCom that produces code documentation using NLP techniques and by combining the lexical and structure information of source code (Hybrid-DeepCom) \cite{hu2018deep, hu2020deep}.

Recent success of pre-trained transformer models in several domains encouraged researchers to also utilize those for automated documentation generation. In fact, they are found to be the state-of the-art performers in this task. CodeBERT, a BERT-based model pre-trained on large scale natural and programming language data, showed great performance in automated documentation generation \cite{feng2020codebert}. Some other transformer-based pre-trained models i.e., PLBART \cite{ahmad2021unified}, CoTexT \cite{phan2021cotext} recently outperformed CodeBERT in the context of documentation generation.

Although pre-trained transformer models have shown promise in code documentation generation, we are aware of no study to evaluate the effectiveness of GPT-3 in this direction. GPT-3 is the third generation Generative Pre-trained Transformer model developed by OpenAI. GPT-3 has 175B parameters and is trained on very large-scale internet data \cite{brown2020language}. It has been found to achieve high performance in different classification and generation tasks \cite{floridi2020gpt, dale2021gpt, chiu2021detecting, chintagunta2021medically}. Similar to other transformer models, GPT-3 architecture can also be used in different software engineering tasks that involve both natural and programming language understanding. In fact, OpenAI released a dedicated model, Codex for this task. Codex is a GPT-3 like model which is trained on large-scale GitHub data. Codex is trained on over a dozen of programming languages like Python, Java, PHP, JavaScript, and so on \cite{chen2021evaluating}. The 
\href{https://www.youtube.com/watch?v=SGUCcjHTmGY}{video demo} of  OpenAI shows that Codex could generate source code from a given requirement specified in Natural Language. Since then researchers have been investigating Codex for the automation of several SE tasks like code generation \cite{finnie2022robots}, code repair \cite{prenner2021automatic}, security bug-fix \cite{pearce2021can}, simulation modeling \cite{jackson2022natural}. The official documentation of Codex mentions that it is also capable of automatic documentation generation. However, we are not aware of any systematic evaluation of Codex to produce code documentation.


In this paper, we conducted a preliminary case study to investigate the effectiveness of Codex in code documentation. We analyzed the code documentation automatically produced by Codex for six programming languages: Python, Java, PHP, GO, JavaScript, Go, and Ruby. To be specific, we evaluated Codex on CodeSearchNet dataset \cite{husain2019codesearchnet} and compared its performance with existing models \cite{feng2020codebert, ahmad2021unified, phan2021cotext, parvez2021retrieval}. Unlike other pre-trained transformer models, GPT-3 often performs well at different downstream tasks without any kind of re-training or fine-tuning. Instead, it is adapted to a task by one (or few)-shot learning where the model is provided with a task description and one (or few) example(s) of that task. In fact, simply providing the task description without any examples (zero-shot learning) often yields good results. In this study, we have experimented Codex with both zero and one-shot learning. We found that Codex with one-shot learning shows state-of-the-art overall performance in documentation generation outperforming all previous models with an average BLEU score of 20.63. As per language specific performance, it outperforms all the models in four out of six languages (i.e., Python, Ruby, JavaScript, GO) with good margin. For the other two languages i.e., Java and PHP, it becomes the second best performer, where it was slightly outperformed by REDCODER and CodeBERT, respectively. We also conducted several qualitative analyses of the Codex documentations in terms of Flesch-Kincaid Grade Level (readability), Documentation Length (quantity), and TF-IDF (informativeness). We found that the generated documentations are close to the actual ones based on these metrics. In fact, we observed that some Codex generated documentation might contain more comprehensible information than the actual ones. We found that even with very basic setup (one-shot learning), Codex is capable of state-of-the-art performance in this field. To the best of our knowledge, ours is the first systematic evaluation of GPT-3 Codex model in automated code documentation generation.

\nd\textbf{Replication Package.} \url{https://github.com/disa-lab/CodeDoc_GPT-3_ASE22}
\section{Related Work}\label{sec:related-work}
Related work on automatic code documentation can be three types: Template, Information retrieval, and Learning-based. 

\nd\textbf{Template-based} uses an automatic tool to insert information according to pre-defined rules and layout. Sridhara et al. utilized natural language templates to capture the key statements from a Java method and build a method level summary \cite{Sridhara-SummaryCommentsJavaClasses-ASE2010}. Rastkar et al. and Moreno et al. used heuristics to extract and summarize information from source code \cite{rastkar2011generating, moreno2013automatic}. Mcburney et al. merged contextual information with method statements \cite{mcburney2014automatic}. Abid et al. produced summary for C++ methods by stereotyping them with their source code analysis framework (srcML) \cite{abid2015using}. Rai et al. summarized Java code that uses code level nano-patterns \cite{rai2017method}.


\nd\textbf{Information Retrieval} approaches like latent semantic indexing (LSI) and vector space modeling (VSM) have been employed by Haiduc et al. to generate documentation for classes and methods \cite{haiduc2010supporting, haiduc2010use}. Eddy et al. extended their work by exploiting a hierarchical topic model \cite{eddy2013evaluating}. Wong et al. employed code clone detection to find similar code snippets from StackOverflow and automatically mined source code descriptions for comment generation \cite{wong2013autocomment, wong2015clocom}.

\nd\textbf{Learning-based} approaches mostly use deep-learning techniques to learn latent features from source-code. Iyer et al. proposed an LSTM-based network, CODE-NN, that was trained on Stack Overflow data to generate C\# and SQL code summaries \cite{iyer2016summarizing}. Allamanis et al. used an attention neural network that employs convolution to learn local, time-invariant features \cite{allamanis2016convolutional}. Barone et al. built a dataset of Python functions and their docstrings using GitHub data and employed Neural Machine Translation (NMT) to generate docstrings from given functions \cite{barone2017parallel}. Wan et al. proposed a reinforcement learning framework that incorporates the abstract syntax tree (AST) along with the sequential code content \cite{wan2018improving}. Hu et al. developed DeepCom and Hybrid-Deepcom to generate code comments by learning from a large corpus and by combining lexical and structural information of code \cite{hu2018deep,hu2020deep}. Chen and Zhou designed a neural framework called BVAE to improve code retrieval and summarization tasks \cite{chen2018neural}. Several studies employed transformer-based models, e.g., Ahmad et al. used self-attention based transformer model \cite{ahmad2020transformer} while Wang et al. developed a BERT-based model called Fret \cite{wang2020fret}. Feng et al. presented CodeBERT, pre-trained on 2.1M bimodal data points (codes and descriptions) and 6.4M unimodal code across six languages (Python, Java, JavaScript, PHP, Ruby, and Go) \cite{feng2020codebert}. They evaluated CodeBERT on two downstream NL-PL tasks (code search and documentation) by fine-tuning model parameters. Gao et al. proposed the concept of Code Structure Guided Transformer for source code summarization that incorporates code structural properties into transformer to improve performance \cite{gao2021code}. Phan et al. presented CoTexT that learns the representative context between natural \& programming language and performs several downstream tasks including code summarization \cite{phan2021cotext}. Ahmad et al. developed PLBART, a sequence-to-sequence model, pre-trained on a large set of Java and Python functions and their textual descriptions collected from Github and
Stack Overflow \cite{ahmad2021unified}. Later, Parvez et al. showed that the addition of relevant codes/summaries retrieved from a database (such as GitHub and Stack Overflow) can improve the quality of documentation produced by a generator model \cite{parvez2021retrieval}.

Though transformer-based models showed promise in documentation generation, GPT-3 model had not been systematically evaluated for this task yet in spite of its success and popularity. Our study employed GPT-3 based Codex model for documentation generation and compared its performance with existing approaches.


\section{Experiment}
\begin{figure}[t]
  \centering
  \includegraphics[scale=.89]{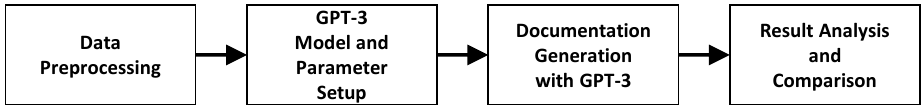}

  \caption{A schematic overview of our study}

  \label{fig:workflow_gpt3_documentation}
\vspace{-1mm}
\end{figure}

In this section, we present the results of our preliminary investigation of the effectiveness of GPT-3 for source code documentation generation. A schematic overview of the major steps of our study is shown at Figure \ref{fig:workflow_gpt3_documentation} and are described below.

\subsection{Dataset Collection and Preprocessing}
\label{subsec:dataset}
To evaluate the effectiveness of our method, we used CodeSearchNet \cite{husain2019codesearchnet}, a widely used
dataset for different downstream SE tasks. This dataset is also included in CodeXGLUE \cite{lu2021codexglue}, a machine learning benchmark dataset for code understanding and generation. It incorporates a large number of code and documentation pairs coming from six different languages i.e., Java, Python, PHP, GO, JavaScript, and Ruby. We applied the same data processing recommended by Feng et al. where they evaluated CodeBERT for documentation generation \cite{feng2020codebert}. We first removed comments from all the codes and then removed examples where i) the codes cannot be parsed into an abstract syntax tree, ii) the number of tokens in the documentation is less than 3 or greater than 256, iii) documentation contains special tokens such as $<$img$>$, https://, etc., iv) the language is not English. The statistics of the clean dataset are available at Table \ref{tab:dataset_stat_codesearchnet}. 

\begin{table}
\centering
\caption{Statistics of CodeSearchNet \cite{husain2019codesearchnet}}
\label{tab:dataset_stat_codesearchnet}
\scalebox{.78}
{
\begin{tabular}{lccc} 
\toprule
\textbf{Language} & \textbf{Train} & \textbf{Valid} & \textbf{Test}  \\ 
\hline
Java              & 164,923        & 5,183          & 10,955         \\
Python            & 251,820        & 13,914         & 14,918         \\
PHP               & 241,241        & 12,982         & 14,014         \\
GO                & 167,288        & 7,325          & 8,122          \\
JavaScript        & 58,025         & 3,885          & 3,291          \\
Ruby              & 24,927         & 1,400          & 1,261          \\
\bottomrule
\end{tabular}
}
\end{table}

\subsection{GPT-3 Model and Parameter Setup}

\textbf{Model.} Generative Pre-trained Transformer-3 (GPT-3) \cite{brown2020language}, developed by \href{https://openai.com/}{OpenAI}, is an autoregressive language model that employs deep learning to generate human-like text. It has 175 billion parameters and 96 layers trained on $\sim$45 TB of text data coming from different web contents (such as Wikipedia). It has promising results in various kinds of NLP tasks e.g., question-answering, summarizing, translation. Not only that, GPT-3 has also shown great success in software engineering tasks like code generation from natural language command. Hence, we felt motivated to evaluate its effectiveness in code documentation generation as well. For our experiment, we used Codex, a descendant of GPT-3, that is trained on both natural language and billions of lines of public code from GitHub. Codex can understand various programming languages such as Python, Java, JavaScript, Go, Perl, PHP, Ruby, etc. Codex has already been used successfully to automate several SE tasks e.g., code generation \cite{finnie2022robots}, code repair \cite{prenner2021automatic}, security bug-fix \cite{pearce2021can}. 

\nd\textbf{Prompt Engineering.} The interaction with GPT-3/Codex takes via prompt engineering, where a task description is provided as the input (prompt) and the model (GPT-3) performs the desired task (generates text) accordingly. There are several ways of prompting with GPT-3 models i.e., \textit{zero-shot, one-shot, few-shot learning}. In \textit{zero-shot learning}, the model is expected to generate an answer without providing any example. In fact, no additional information other than the task description itself is given in the prompt. On the other hand, \textit{one-shot} and \textit{few-shot learning} involve giving one (i.e., one-shot) or more than one (i.e., few-shot) examples in the prompt, respectively. In this study, we have experimented with \textit{zero-shot} and \textit{one-shot learning}. In zero-shot learning, we just tell the model to generate a documentation for a given source code in the prompt. In one-shot learning, we randomly select one sample (i.e., code-documentation pair) from the train set of the corresponding language (from CodeSearchNet), provide it in the prompt (as example) and then ask the model to generate a documentation for another source code by learning from the provided example of code-documentation pair. Figure \ref{fig:one_shot_prompt} depicts the one-shot prompt format we used for documentation generation. 

\begin{figure}[t]
\centering
    \begin{lstlisting}[frame=single]
`Code:`
    def add(x, y):
        return x+y
`Documentation:` Adds two numbers.
`Code:`
    def subtract(x, y):
        return x-y
`Documentation:` ?[To be generated by Codex]?
    \end{lstlisting}
    
\caption{Sample prompt format for one-shot learning}
    \label{fig:one_shot_prompt}    
\end{figure}

\nd\textbf{Parameter Settings.}
There are a number of parameters involved with GPT-3 based models. One such parameter is \textit{Temperature} that controls the randomness of the generated output (range 0 to 1). Another randomness parameter is \textit{Top-p} (range 0 to 1) that controls how unlikely words can get removed from the sampling pool by choosing from the smallest possible set (of words) whose cumulative probability exceeds \textit{p}. As recommended by \href{https://beta.openai.com/docs/guides/code/best-practices}{Codex official documentation}, we set the temperature at a low value (0.2) while keeping top-P at its default value (1.0). We also keep \textit{Frequency} and \textit{Presence penalties} at their default values (0.0) which control the level of word repetition in the generated text by penalizing them based on their existing frequency and presence. We use the \textit{max tokens size} as 256 since our formatted dataset does not contain any larger ($>$256) documentation (see Section \ref{subsec:dataset}).


\subsection{Evaluation of Generated Documentation}
Since GPT-3 models are currently subject to response limit and cost, we used a statistically significant sample size for this study instead of using the whole test sets for different languages. As depicted in Table \ref{tab:dataset_stat_codesearchnet}, the largest test set in CodeSearchNet belongs to Python consisting of 14,918 samples and a statistically significant sample size from that would be 375 with 95\% confidence interval and 5\% error margin. However, we randomly selected 1000 ($>$375) samples from each test set (i.e., total 6 sets of 1K samples for 6 languages) and evaluated Codex model on them. For evaluation, we used BLEU score \cite{papineni2002bleu}, a popular evaluation metric for machine-generated text that calculates the n-gram similarity of a generated and reference text. Since the generated documentation can be short at times and higher order n-gram might not overlap, we used smoothed BLEU score \cite{lin2004orange} as recommended by prior works \cite{feng2020codebert, ahmad2021unified, parvez2021retrieval}.  


\subsubsection{Performance Analysis}
Table \ref{tab:result_on_documentation_generation} compares Codex's performance with several SOTA models for documentation generation: Seq2Seq \cite{sutskever2014sequence}, Transformer \cite{vaswani2017attention}, RoBERTa \cite{liu2019roberta}, CodeBERT \cite{feng2020codebert}, PLBART \cite{ahmad2021unified}, CoTexT \cite{phan2021cotext}, REDCODER \cite{parvez2021retrieval}. We observed that though Codex with zero-shot learning could not achieve satisfactory results (mostly because it fails to learn the expected documentation format), the performance greatly improves with one-shot learning. In fact, Codex (with one-shot) shows the best overall performance among all approaches with an average BLEU score of 20.63 while the nearest competitor CoTexT achieves 18.55 (11.21\% improvement). In language specific performance, it significantly outperforms other models in all languages except two i.e., Java and PHP. In Java and PHP, Codex achieves BLEU scores of 22.81 and 25.13, which are slightly outperformed by REDCODER (22.95) and CodeBERT (25.16) respectively. Here, REDCODER is not an individual model on its own, rather it is a retrieval approach that can be used with other generative models to enhance their performances. In the original paper \cite{parvez2021retrieval}, the authors used PLBART \cite{ahmad2021unified} as the base model for REDCODER where they retrieved relevant summaries from StackOverflow, GitHub, etc. and provided them with the input of PLBART to enhance its performance. Hence, it has some additional overhead (time and resource) while Codex is an all-in-all model itself. Moreover, Codex can perform even better if used with such additional retrieval approach (i.e., REDCODER). On the other hand, CodeBERT and most other reported models had been (re)trained or fine-tuned on task and language-specific datasets while Codex was provided with only zero or one example in our evaluation.    

\begin{table}
\caption{Results on documentation generation (BLEU score)}
\label{tab:result_on_documentation_generation}
\begin{center}
\resizebox{\columnwidth}{!}
{
\begin{tabular}{lccccccc} 
\toprule
\textbf{Model} & \textbf{Ruby} & \textbf{JavaScript} & \textbf{GO} & \textbf{Python} & \textbf{Java} & \textbf{PHP} & \textbf{Overall}  \\ 
\midrule
Seq2Seq \cite{sutskever2014sequence}        & 9.64          & 10.21               & 13.98       & 15.93           & 15.09         & 21.08        & 14.32             \\
Transformer \cite{vaswani2017attention}    & 11.18         & 11.59               & 16.38       & 15.81           & 16.26         & 22.12        & 15.56             \\
RoBERTa \cite{liu2019roberta}        & 11.17         & 11.90               & 17.72       & 18.14           & 16.47         & 24.02        & 16.57             \\
CodeBERT \cite{feng2020codebert}       & 12.16         & 14.90               & 18.07       & 19.06           & 17.65         & 25.16        & 17.83             \\
PLBART \cite{ahmad2021unified}         & 14.11         & 15.56               & 18.91       & 19.30           & 18.45         & 23.58        & 18.32             \\
CoTexT (2-CC) \cite{phan2021cotext}  & 13.07         & 14.77               & 19.37       & 19.52           & 19.1          & 24.47        & 18.38             \\
CoTexT (1-CC) \cite{phan2021cotext}  & 14.02         & 14.96               & 18.86       & 19.73           & 19.06         & 24.58        & 18.55             \\
REDCODER \cite{parvez2021retrieval}       & -             & -                   & -           & 21.01           & 22.94         & -            & N/A               \\
REDCODER-EXT \cite{parvez2021retrieval}   & -             & -                   & -           & 20.91           & 22.95         & -            & N/A               \\ 
\midrule
Codex (0-shot) &      5.41         &      9.83               & 15.80            &       18.93          &        13.59       & 13.32             &           12.81        \\
Codex (1-shot) &      16.04         &           16.58          &    20.94         &      22.28           &       22.81        &        25.13      &   20.63                \\
\bottomrule
\end{tabular}
}
\end{center}
\end{table}




\subsubsection{Qualitative Analysis}
\begin{figure}[t]
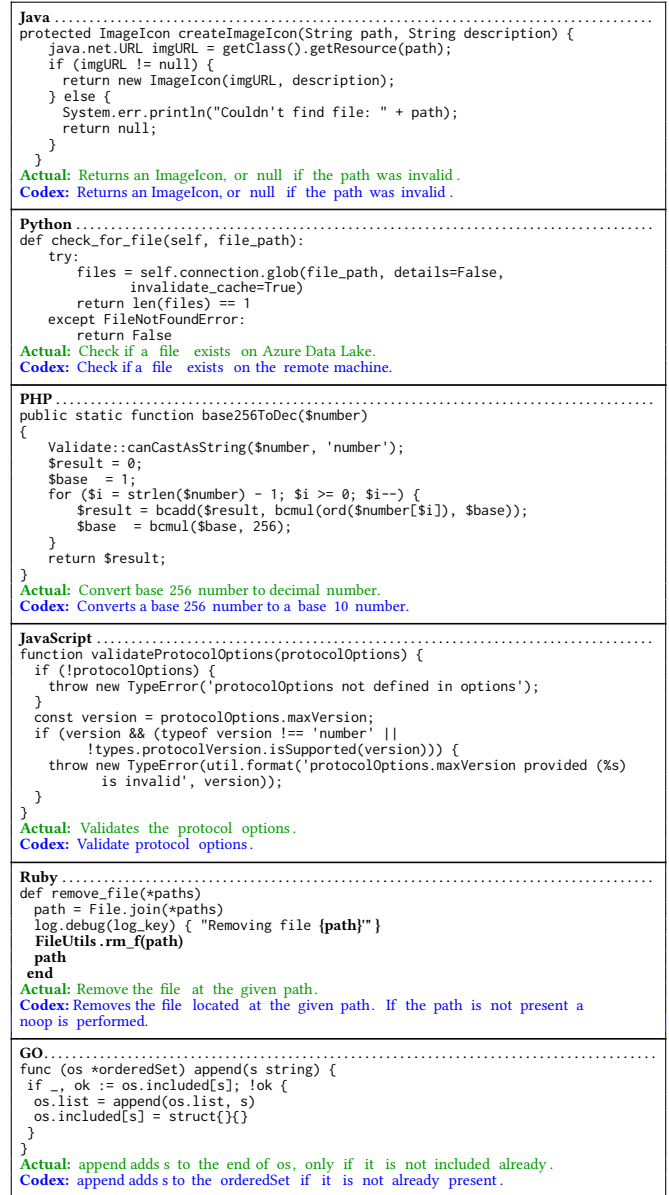

    \begin{lstlisting}[frame=single]
`Java`//
protected ImageIcon createImageIcon(String path, String description) {
    java.net.URL imgURL = getClass().getResource(path);
    if (imgURL != null) {
      return new ImageIcon(imgURL, description);
    } else {
      System.err.println("Couldn't find file: " + path);
      return null;
    }
  }
@Actual:@ ^Returns an ImageIcon, or null if the path was invalid.^
~Codex:~ ?Returns an ImageIcon, or null if the path was invalid.?
    \end{lstlisting}
    
    \begin{lstlisting}[frame=single]
`Python`//
def check_for_file(self, file_path):
    try:
        files = self.connection.glob(file_path, details=False, invalidate_cache=True)
        return len(files) == 1
    except FileNotFoundError:
        return False
@Actual:@ ^Check if a file exists on Azure Data Lake.^
~Codex:~ ?Check if a file exists on the remote machine.?
    \end{lstlisting}
    
    \begin{lstlisting}[frame=single]
`PHP`//
public static function base256ToDec($number)
{
    Validate::canCastAsString($number, 'number');
    $result = 0;
    $base   = 1;
    for ($i = strlen($number) - 1; $i >= 0; $i--) {
        $result = bcadd($result, bcmul(ord($number[$i]), $base));
        $base   = bcmul($base, 256);
    }
    return $result;
}
@Actual:@ ^Convert base 256 number to decimal number.^
~Codex:~ ?Converts a base 256 number to a base 10 number.?
    \end{lstlisting}
    
    \begin{lstlisting}[frame=single]
`JavaScript`//
function validateProtocolOptions(protocolOptions) {
  if (!protocolOptions) {
    throw new TypeError('protocolOptions not defined in options');
  }
  const version = protocolOptions.maxVersion;
  if (version && (typeof version !== 'number' || !types.protocolVersion.isSupported(version))) {
    throw new TypeError(util.format('protocolOptions.maxVersion provided (%s) is invalid', version));
  }
}
@Actual:@ ^Validates the protocol options.^
~Codex:~ ?Validate protocol options.?
    \end{lstlisting}
    
    \begin{lstlisting}[frame=single]
`Ruby`//
def remove_file(*paths)
  path = File.join(*paths)
  log.debug(log_key) { "Removing file `{path}'" }
  FileUtils.rm_f(path)
  path
 end
@Actual:@ ^Remove the file at the given path.^
~Codex:~ ?Removes the file located at the given path. If the path is not present a
noop is performed.?
    \end{lstlisting}
    
    \begin{lstlisting}[frame=single]
`GO`//
func (os *orderedSet) append(s string) {
	if _, ok := os.included[s]; !ok {
		os.list = append(os.list, s)
		os.included[s] = struct{}{}
	}
}
@Actual:@ ^append adds s to the end of os, only if it is not included already.^
~Codex:~ ?append adds s to the orderedSet if it is not already present.?
    \end{lstlisting}
    
\caption{Examples of documentation by Codex (1-shot)}
    \label{fig:codex_doc_examples}    
\end{figure}

As suggested by Schreck et al. \cite{schreck2007documentation}, we used Documentation Length and Flesch-Kincaid Grade
Level \cite{dubay2004principles} to measure the Quantity and Readability of the generated documentation. We find that the average Flesch-Kincaid score of the Codex generated documentations is 5.97 with an average length of 8 words (per documentation) while the average Flesch-Kincaid score of the actual documentations is 6.77 with an average length of 12 words. Hence, the generated documentations are close to the actual ones in terms of quantity and readability. We further analyzed the informativeness of the generated documentation with respect to the actual ones using TF-IDF. To calculate the TF-IDF of a particular documentation, we add TF-IDF scores of all the words of the documentation (except stop-words). The average TF-IDF of Codex generated documentations is 1.94 while for actual documentations it is 2.28. Hence, the informativeness of the generated documentation is satisfactory in terms of comparative TF-IDF with the actual ones. We show some examples of Codex generated documentations in Figure \ref{fig:codex_doc_examples}. We see that in some cases, Codex can generate more comprehensible or additional information compared to the actual ones. In the example of PHP mentioned in Figure \ref{fig:codex_doc_examples}, Codex mentions ``base-10 number" instead of ``decimal'' which is more understandable for users. In the example of Ruby, Codex provides an extra information ``If the path is not present a noop is performed." which is correct but was not present in the actual documentation.

\section{Conclusion and Future Work}

We explored GPT-3 Codex for automatic documentation generation and compared its performance with existing approaches. While previous approaches are subject to task/language-specific retraining or fine-tuning, Codex shows SOTA performance even with very basic settings. In future, we intend to investigate Codex more in-depth in terms of parameter tuning, few shot learning, fine-tuning to improve its performance even further. We also plan to employ Codex to fix documentation issues (e.g., doc smells) that we found in our earlier study \cite{khan2021automatic}. In particular, 
this study has a number of limitations that we want to address in the future. First, we used 1K samples for each languages to evaluate Codex generated documentation. Though the sample size is statistically significant, we intend to test Codex on more samples in the future. Second, we limited our investigation only to zero and one-shot learning. We will extend the investigation by also analyzing few-shot learning. Third, we randomly picked one sample from the corresponding train set to use it in the one-shot learning. However, using different samples in one-shot learning might yield different outcome which has not been explored. Fourth, like other pre-trained transformer models, GPT-3 also supports fine-tuning. However, latest versions of GPT-3 models (e.g., text-davinci, Codex) are not available for fine-tuning. We could fine-tune GPT-3 when it is available to do so. Finally, systematic investigation of the effect of different parameters associated with GPT-3 has not been done in this study. We picked the parameter values  based on the official documentation.

\nd\textbf{Acknowledgement.} This work was funded by Natural Sciences and Engineering Research Council of Canada, University of Calgary, Alberta Innovates, and Alberta Graduate Excellence Scholarship.




\bibliographystyle{ACM-Reference-Format}
\bibliography{consolidated}

\end{document}